\documentclass[]{aastex631}

\usepackage{graphicx}
\usepackage{amsmath}
\usepackage{amsfonts}
\usepackage{wasysym}
\usepackage{breqn}
\usepackage{comment}
\newcommand{\change}[1]{{#1}}

\begin{document}

\title{Stable stratification of the helium rain layer yields vastly different interiors and magnetic fields for Jupiter and Saturn}

\author[0009-0005-5613-3026]{S. Markham}
\affiliation{New Mexico State University, Dept. of Astronomy, Las Cruces, NM 88003}
\affiliation{Universit\'{e} C\^{o}te d'Azur, Observatoire de la C\^{o}te d'Azur, CNRS, Laboratoire Lagrange, Nice, France}
\email{markham@nmsu.edu}

\author[0000-0002-7188-8428]{T. Guillot}
\affiliation{Universit\'{e} C\^{o}te d'Azur, Observatoire de la C\^{o}te d'Azur, CNRS, Laboratoire Lagrange, Nice, France}


\begin{abstract}

At sufficiently high pressures ($\sim$Mbar) and low temperatures ($\sim 10^3-10^4$K), hydrogen and helium become partly immiscible. Interpretations of Jupiter and Saturn's magnetic fields favor the existence of a statically stable layer near the Mbar pressure level. From experimental and computational data for the hydrogen-helium phase diagram we find that moist convection and diffusive convection are inhibited, implying a stable helium rain layer in both Jupiter and Saturn.
However, we find a significant difference in terms of structure and evolution: In Jupiter, helium settling leads to a stable yet super-adiabatic temperature gradient that is limited by conductive heat transport. The phase separation region should extend only a few tens of kilometers instead of thousands in current-day models, and be characterized by a sharp increase of the temperature of about 500K for standard phase separation diagrams. In Saturn, helium rains occurs much deeper, implying a larger helium flux relative to planetary mass. We find that the significant latent heat associated with helium condensation implies that a large fraction, perhaps close to 100\%, of the planet's intrinsic heat flux may be locally transported by the sinking helium droplets. This implies that Saturn may possess a much more extended helium-rain region.
This also accounts, at least qualitatively, for the differences in strength and characteristics of the magnetic fields of the two planets. Dedicated models of magnetic field generation in both planets may offer observational constraints to further refine these findings.

\end{abstract}

\section{Introduction}
\label{sec:intro}
Hydrogen and helium coexist in any proportion as molecular gases, but become immiscible at high pressures relevant to the interiors of Jovian planets \citep{salpeter1973, stevenson-salpeter1977a}.
Helium settling has been suggested as a heat source to explain Saturn's observed luminosity, which is not consistent with an adiabatic, homogeneous evolution \citep{stevenson-salpeter1977b, fortney+hubbard2003, mankovich-fortney2020}. 
The first-principles prediction of a hydrogen-helium miscibility gap have recently been convincingly demonstrated computationally \citep{morales+2013, schottler-redmer2018} and have gained credible claims of laboratory evidence \citep{brygoo+2021}. \\

The dynamical nature of the hydrogen-helium phase separating region is a subject of debate. 
\cite{stevenson-salpeter1977b} considers the general case of a simultaneous gradient in composition and temperature applied to hydrogen-helium demixing, and acknowledges a range of possibilities including convection, diffusive convection, or stable stratification depending on the gradients and transport properties. 
The general problem has received much attention in the context of giant planetary interiors, primarily in the context of understanding the heavy-element distribution, predicting layered semi-convection driven by overstable double-diffusive motions \citep[e.g.,][]{leconte-chabrier2012, nettelmann++2015, mankovich+2016, fuentes+2022}. 
However, observations of Jupiter and Saturn appear to favor a stably stratified helium-rain layer. 
Saturn's seismic spectrum inferred from ring seismology favors the existence of a deep, extended layer of static stability that may evince a stable helium rain layer \citep{mankovich-fuller2021}. 
Saturn's remarkably axisymmetric magnetic field aligned along its spin-axis can be explained using a conductive layer of static stability above the dynamo region \citep{stevenson1980, stevenson1982, stanley2010, cao+2011}. 
Jupiter's magnetic field as observed by Juno likewise has been interpreted to include a layer of static stability near the expected hydrogen-helium demixing level. 
In particular, by fitting the attenuation slope of higher order harmonics of Jupiter's magnetic field, these studies find Jupiter's magnetic field Lowes radius to be 80--83\% of Jupiter's radius \citep{connerney+2022, sharan+2022}. 
Jupiter's Lowes radius corresponds to a pressure level of {$\sim$1.2 to 2\,Mbar}, near where we expect hydrogen-helium phase separation to occur. 
On the Earth, the Lowes radius corresponds to the top of the dynamo-generating liquid outer core, and these works thus interpret Jupiter's Lowes radius to likely correspond to the top of Jupiter's dynamo-generating region as well. 
Numerical dynamo models of Jupiter's magnetic field likewise favor the existence of a statically stable layer, although the best-fit models place the stable layer at a shallower depth between 85--95\% of Jupiter's radius \citep{moore+2018, tsang-jones2020, moore+2022}. 
While possible stable stratification due to helium settling has been suggested by all these works in order to explain Jupiter's magnetic field, the stability of the helium rain layer has not yet been rigorously established by theory. \\

In this work, we posit a theoretical rationale for the stability of the helium rain layer against both ordinary and diffusive convection on the basis of convective inhibition by phase separation. 
Super-adiabatic temperature gradients may remain stable against convection if stabilized by a corresponding compositional gradient \citep{ledoux1947}. 
Convective inhibition refers to cases when the compositional gradient is itself a function of the temperature gradient, for example on a multi-component phase boundary. 
If the density difference between both components is sufficient, and the composition is a sensitive enough function of temperature, then some mixtures under certain thermodynamic conditions can be stable against convection even subject to arbitrarily steep temperature gradients. 
The notion of convective inhibition by phase transitions has been studied in some detail in the context of volatile condensation in hydrogen atmospheres \citep{guillot1995, leconte+2017, friedson-gonzales2017, markham-stevenson2021}. 
The principle of convective inhibition, heretofore based on linear theory, has recently been corroborated by computational simulations that observe convective inhibition as an emergent phenomenon from physics-based hydrodynamic models under appropriate conditions \citep{ge++2023, leconte++2024}. 
Convective inhibition has also been shown to apply outside of the ideal gas atmospheric context \citep{markham+2022} and can be generalized to many multi-component systems undergoing phase separations; the phenomenon has also been applied to equilibrium chemical reactions \citep{misener++2023}. 
\cite{schottler-redmer2018} has carried out a full characterization of the three dimensional hydrogen-helium phase diagram (that is, varying both thermodynamic variables and composition) using ab initio density functional theory (DFT). 
Combining such a phase diagram with a suitable equation of state (EoS) allows the possibility of assessing the applicability of convective inhibition to hydrogen-helium phase separation. 
Such is the aim of this work. \\

The main text focuses on highlighting our primary methods and findings intended to be useful for modelers of internal structures and thermal evolution of Jovian planets. 
In Sec.~\ref{sec:stability}, we introduce the concept of convective inhibition in the context of the hydrogen-helium phase separating region of Jovian planets, and quantify the convective inhibition criterion. 
Next in Sec.~\ref{sec:structure} we comment on the effect of convective inhibition in the helium rain layer on the structure, evolution, and magnetic fields of Jovian planets. 
Finally we highlight our most important findings in Sec.~\ref{sec:conclusion}. 
The Appendix includes further details that may be of interest to theorists: 
we consider the extent to which realistic transformations of the controversial phase diagram may affect our findings, discuss the non-uniqueness of phase diagram intersection points, and derive the effect of latent heat on the pseudo-adiabatic lapse rates of phase separating mixtures under non-ideal conditions.

\section{Stability of the helium rain layer}
\label{sec:stability}
\subsection{The effective coefficient of thermal expansion}
\label{sec:alpha}
A zeroth-order theory of convection can be summarized using the layman's aphorism ``heat rises.'' 
This theory is approximately correct for most low-viscosity fluids. 
Most materials have a positive coefficient of thermal expansion
\begin{equation}
\alpha \equiv \rho \left( \frac{\partial (1/\rho)}{\partial T} \right)_{p,x}, 
\label{eq:alpha}
\end{equation}
where $1/\rho$ is the specific volume, and $p$ and $x$ refer to pressure and composition respectively. 
For an inviscid substance of uniform composition in a gravitational field aligned in the $\hat{\mathbf{z}}$ direction, the condition
\begin{equation}
\alpha \left[ \frac{dT}{dz} - \left(\frac{dT}{dz}\right)_{\rm ad} \right] < 0
\label{eq:instability-criterion}
\end{equation}
produces a Rayleigh-Taylor instability, so that super-adiabatic (i.e., $-dT/dz > -(dT/dz)_{\rm ad}$) temperature gradients cannot persist in nature unless $\alpha < 0$. 
The Appendix~\ref{sec:pseudo} discusses the appropriate form of $\left(\frac{dT}{dz}\right)_{\rm ad}$ for a non-ideal phase separating pseudo-adiabat relevant to the helium rain region.
\\

\begin{figure}
\centering
\includegraphics[width=0.5\textwidth]{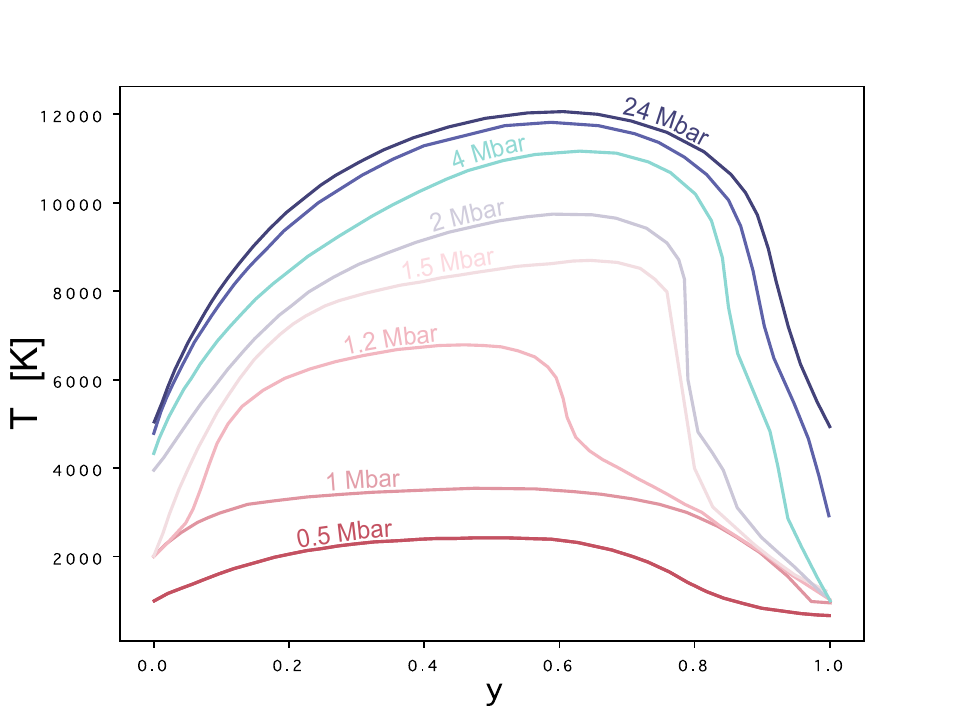}
\caption{Coexistence curves for a hydrogen-helium mixture adapted from \cite{schottler-redmer2018} with a temperature translation $\delta T = 650$K following \cite{mankovich-fortney2020} (for further discussion, see Appendix~\ref{sec:phase-diagram}). 
At each pressure level for a given temperature below the peak of the curve, there exist two coexisting phases of different composition, one helium-rich and one helium-poor.}
\label{fig:coexistence}
\end{figure}

In the case of a phase-separating mixture, we can define an ``effective'' coefficient of thermal expansion $\alpha_{\rm eff}$ that replaces $\alpha$ in Eq.~\ref{eq:instability-criterion}. 
The effective coefficient of thermal expansion accounts for the possibility that the composition of a working substance can change as its temperature changes, for example as helium-enriched raindrops settle out when a mixture cools on the boundary of the hydrogen-helium miscibility gap. 
The effective coefficient of thermal expansion is \citep{markham+2022}
\begin{equation}
\alpha_{\rm eff} = \alpha + \rho \left( \frac{\partial y}{\partial T} \right)_{\rm ph} \left(\frac{\partial (1/\rho)}{\partial y}\right)_{p,T},
\label{eq:alpha_eff}
\end{equation}
where $y$ is the molar mixing ratio of helium. 
$\left( \frac{\partial y}{\partial T} \right)_{\rm ph}$ refers to evaluating the composition along the coexistence phase boundary from Figure~\ref{fig:coexistence}. 
Eq.~\ref{eq:alpha_eff} assumes thermodynamic equilibrium and ignores transport; this assumption will be challenged in Sect.~\ref{sec:condensation}. 
From Eq.~\ref{eq:alpha_eff}, we see a suitable EoS and phase diagram fully define $\alpha_{\rm eff}$. 
For example, using the ideal gas equation of state and an equilibrium vapor pressure defined by an Arrhenius relationship $p_{\rm saturation} \propto \exp\left(\frac{-\lambda}{R T}\right)$ where $\lambda$ is the specific latent heat and $R$ is the specific gas constant, the substitution of Eq.~\ref{eq:alpha_eff} into $\alpha$ from Eq.~\ref{eq:instability-criterion} exactly reproduces the convective inhibition criterion for moist convection from \cite{guillot1995}. 
However, this more general notation allows us to apply the criterion to non-ideal mixtures. \\

\begin{figure}
\centering
\includegraphics[width=0.7\linewidth]{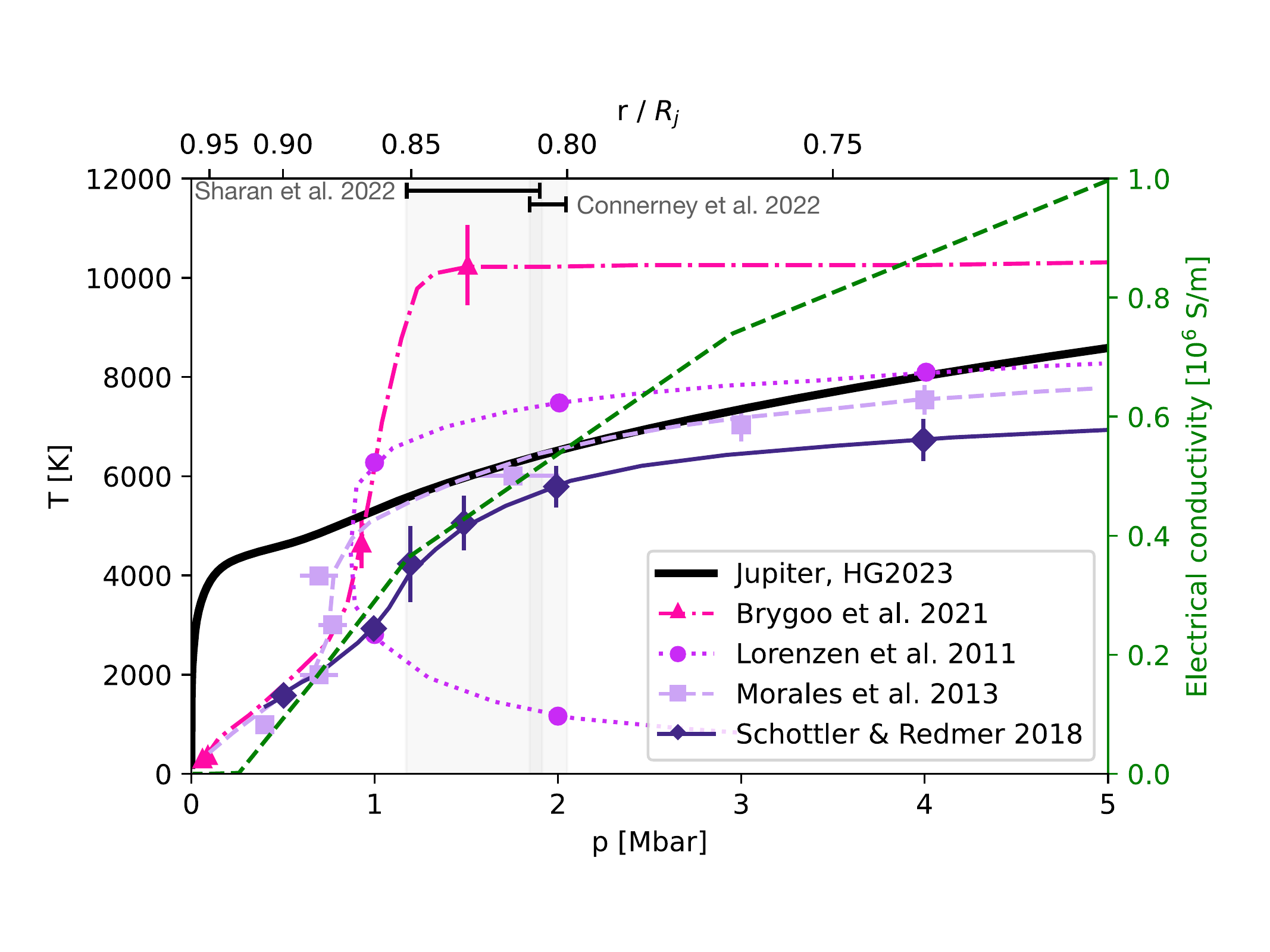}
\caption{A summary of the outstanding problem with respect to the hydrogen-helium phase diagram. 
We show predicted phase boundaries based on four different studies \citep{lorenzen+2011, morales+2013, schottler-redmer2018, brygoo+2021}. 
We compare this to a sample interior structure of Jupiter \citep{howard+2023}, and the Lowes radius estimate error bars with shadows \citep{sharan+2022, connerney+2022}. The green dashed curve, measured on the right hand axis, is a tabulated electrical conductivity profile for Jupiter's interior \citep{french+2012}. }
\label{fig:problem}
\end{figure}

First one must choose a suitable phase diagram for a H-He mixture. 
This turns out to be a non-trivial choice, as there exist a variety of credible published phase diagrams with notably different predictions as shown in Fig.~\ref{fig:problem}. 
We chose to use ab initio DFT simulations from \cite{schottler-redmer2018} for the density of information in pressure, temperature, and composition space. 
Following \cite{mankovich-fortney2020}, we impose an arbitrary $\delta T = 650$K to account for Jupiter's measured helium depletion. 
We show the results on Fig.~\ref{fig:coexistence} \citep[cf. Fig. 2 from][]{schottler-redmer2018}. 
For the EoS, we follow \cite{howard-guillot2023}, i.e., we use the pure EOSs calculated for hydrogen and helium by \cite{chabrier+2019} and account for non-ideal mixing terms assumed proportional to $XY$, the product of the mass mixing ratios of hydrogen and helium by using the hydrogen-helium EOS from \cite{militzer+hubbard2013ApJ} \citep[see also][]{chabrier+debras2021}.
\begin{figure*}
\centering
\includegraphics[width=.75\textwidth]{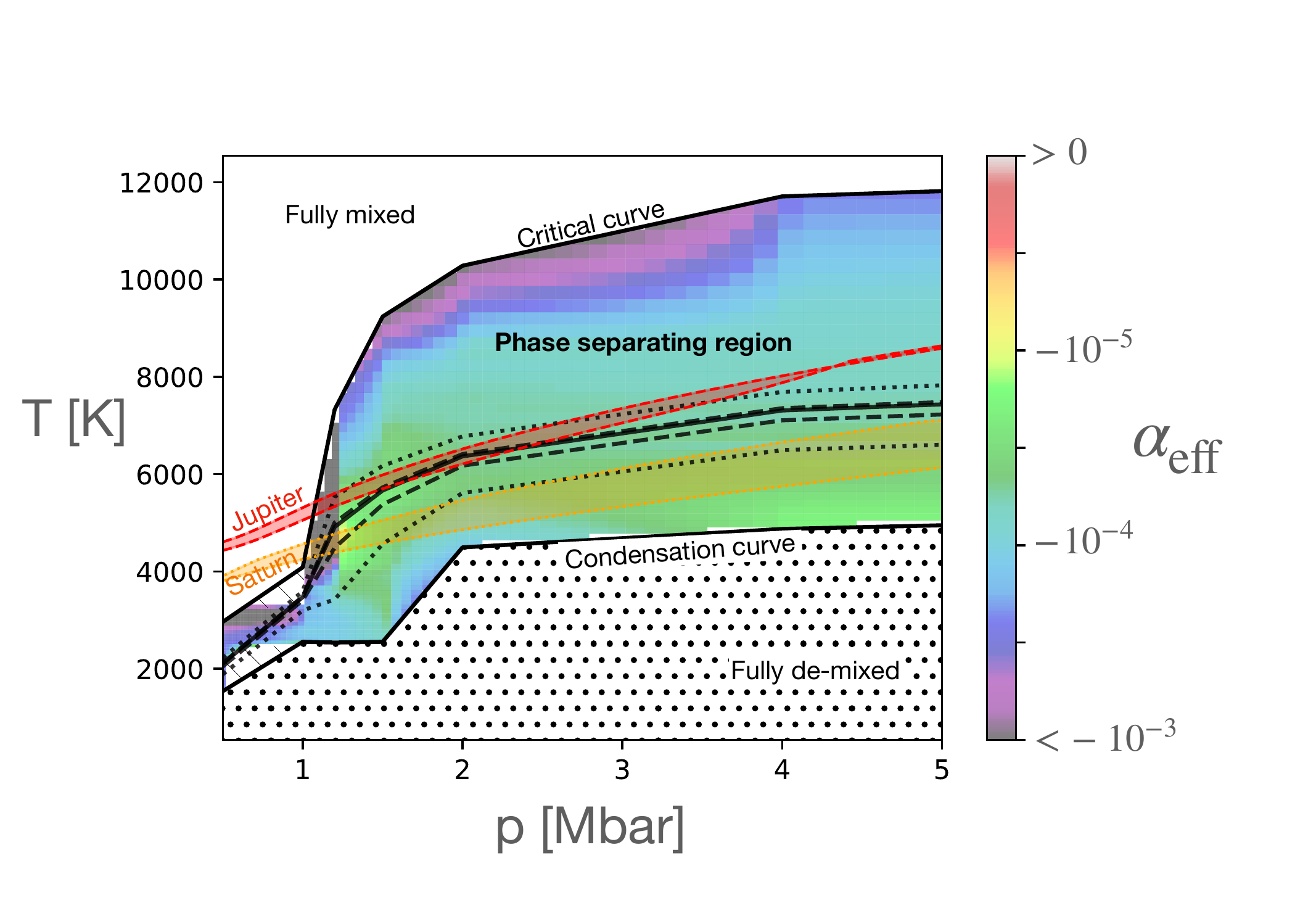}
\caption{Effective coefficient of thermal expansion $\alpha_{\rm eff}$ from Equation~\ref{eq:alpha_eff} in pressure, temperature space. 
Values calculated using the EoS from \cite{howard-guillot2023} and by interpolating the phase boundary surface in pressure--temperature--composition space from Fig.~\ref{fig:coexistence}. 
The red region labeled Jupiter is bounded between Jupiter interior models from \cite{militzer+2022} (cooler) and  \cite{howard+2023} (warmer).
Similarly the orange region labeled Saturn is bounded between Saturn interior models from \cite{mankovich-fortney2020} (cooler) and \cite{hubbard+2009} (warmer). 
We found $\alpha_{\rm eff}$ to be negative almost everywhere where hydrogen and helium are phase-separating. 
The phase separating region includes phase boundary curves for various sample compositions, showing y=[5, 7.88, 8.87, 9.1, 15]\% in ascending order. 
}
\label{fig:master}
\end{figure*}
Fig.~\ref{fig:master} shows the hydrogen-helium de-mixing region in pressure-temperature space. 
The colors in Fig.~\ref{fig:master} show the values of the $\alpha_{\rm eff}$, negative nearly everywhere of interest for the helium rain level. 
The colored region is bounded from above by the critical curve, above which hydrogen and helium can mix in all proportions as a supercritical fluid. 
It is bounded from below by what we label the condensation curve, below which hydrogen exists as a pure substance unpolluted by helium. 
Each point in pressure-temperature space implies an equilibrium composition for the hydrogen-rich phase according to the left half of Fig.~\ref{fig:coexistence}. 
We therefore include curves of constant mixing ratio within the demixing region: the solid curve is the solar composition $y=8.87\%$, while the dashed curves refer to the composition of the top and bottom of the stable layer in Jupiter, around $y=7.88\%$ and $y=9.1\%$ based on Galileo measurements of Jupiter's atmosphere, assumed representative of the helium-depleted envelope, and a corresponding helium-enriched interior assuming mass balance of an initially protosolar helium abundance. 
Finally the dotted curves refer to notionally plausible values for Saturn, around $y=5\%$ and $y=15\%$. \\

The key takeaway from Fig.~\ref{fig:master} is that $\alpha_{\rm eff} < 0$ for most relevant parameters. 
\emph{This implies that hydrogen-helium phase separation in thermodynamic equilibrium should be expected to be stable against convection, even under super-adiabatic temperature gradients. } 

\subsection{Stability against diffusive convection}
\label{sec:condensation}
In Sect.~\ref{sec:alpha}, we assumed thermodynamic equilibrium when evaluating the density of a hydrogen-helium mixture as a function of temperature and pressure using a suitable EoS. 
However, in real systems thermodynamic equilibrium will not be reached instantaneously. 
A cooling mixture susceptible to phase separation will separate phases on a finite condensation timescale governed by kinetics. 
Once condensate drops form, they will settle out on a finite timescale governed by fluid dynamics. 
On the other hand, material that is warming up may thermodynamically permit additional helium to dissolve into the mixture, but whether that helium is actually available depends on the local compositional structure and efficiency of diffusion. 
These issues have been studied in detail in the context of planetary atmospheres \citep{leconte+2017, friedson-gonzales2017}. 
\cite{leconte+2017} considered the effect of the condensation timescale $\tau_c$, essentially the time it takes for the system to reach thermodynamic equilibrium. 
They found that ordinary and diffusive convection are suppressed when $\tau_c N < 3$ where $N$ is the Br\"{u}nt V\"{a}is\"{a}l\"{a} frequency. 
\cite{mankovich+2016} estimates $\tau_c \sim 10^{-1}$\,s, so following these numbers we can expect the helium rain layer to be stable against convection so long as $N<30\rm\,s^{-1}$. 
In Section~\ref{sec:conductive}, we find an upper bound for $N$ based on conductive equilibrium to be $N \lesssim 10^{-3}\rm\,s^{-1}$. 
Although the precise numbers may possess some uncertainties, we find $\tau_c N \sim 10^{-4} \ll 3$. 
With more than four orders of magnitude worth of leeway, we conclude that the condensation timescale is not a serious challenge to our central conclusion from Sec.~\ref{sec:alpha}. 
Nevertheless, future work investigating the nucleation process of helium raindrops in more detail is warranted. \\

Complementarily, \cite{friedson-gonzales2017} assumed instantaneous condensation, considering instead the effect of finite diffusivity. 
They found that convection is suppressed for all wavelengths irrespective of fast or slow sedimentation provided that the Prandtl number is sufficiently large, but that unstable modes can exist for slow sedimentation and small Prandtl number. 
We estimate the Prandtl number to be Pr~$\sim 10^{-2}$ \citep{stevenson-salpeter1977b}, sufficiently small that the sedimentation timescale may be important. 
\cite{friedson-gonzales2017} considered the limiting cases where condensed droplets rain out instantly (as is generally assumed, for example, when deriving the moist pseudo-adiabat), or remain suspended indefinitely. 
They found that convection is suppressed in all cases as long as the settling velocities are large compared to buoyant velocities, a condition we expect to hold for helium rain based on arguments from the previous paragraph and from prior works \citep{stevenson-salpeter1977a, mankovich+2016}. 
We therefore do not expect the finite diffusivity in Jupiter's helium rain layer to affect our result that the helium rain layer should be stable, although again further study is warranted.

\section{Structure of the helium rain layer}
\label{sec:structure}
Having argued for the stability of the helium rain layer in Sec.~\ref{sec:stability}, we now describe the expected structure of a stable helium rain layer. 
We begin by placing a lower bound on the stable layer thickness in Sec.~\ref{sec:conductive} under the assumption of conductive-convective equilibrium. 
We then comment on the role of the latent heat of helium rain drops in modifying this prediction in Sec.~\ref{sec:latent} and suggest a possible distinction between the helium rain layers on Jupiter and Saturn. 
Finally in Sec.~\ref{sec:magnetic} we motivate closer inspection of Jupiter and Saturn's magnetic fields as a possible method to place observational constraints on stable layer thickness in future work.

\subsection{Conductive equilibrium as a lower bound for stable layer thickness}
\label{sec:conductive}
\begin{figure}
\centering
\includegraphics[width=\textwidth]{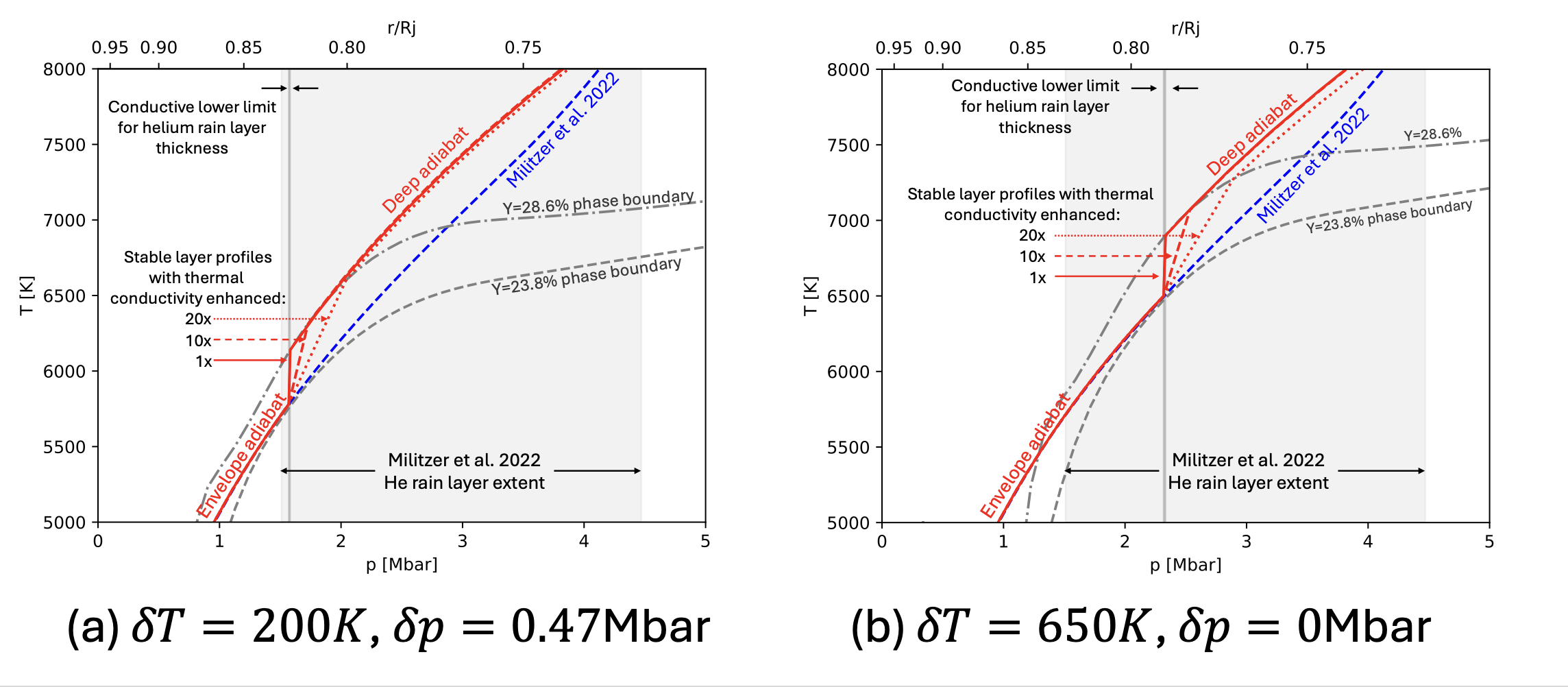}     
\caption{\change{Sample PT structures of the stable layer as conceptualized by this work, compared to \cite{militzer+2022} (blue dashed curve). 
We demonstrate how changes to the phase diagram affect the placement of the helium rain layer by translating phase boundary curves from \cite{schottler-redmer2018} by some offsets in temperature, $\delta T$ and pressure $\delta p$ (see Appendix~\ref{sec:phase-diagram} for details). }
The red curves show the temperature structure envisioned by this work, with the solid red curve assuming a thermal conductivity of 300W/(m~K)
, and for thermal conductivity enhanced by a factor of 10x and 20x for the red dashed and red dotted curves respectively. 
\change{We argue that the top and bottom of the stable helium rain layer should intersect phase boundaries of corresponding composition. }}
\label{fig:lapses}
\end{figure}

If the helium rain layer is stable against convection even subject to arbitrarily steep super-adiabatic temperature gradients, then what is the maximum physically possible temperature gradient?
While possible fluid dynamical processes (for example, turbulent diffusion excited by breaking gravity waves) may transport heat more efficiently, the details of such fluid dynamical phenomena in an environment as poorly understood as Jovian planets' deep interiors would be highly speculative and uncertain. 
Latent heat may also play an important role in regulating energy balance, a topic discussed in some detail in Sec.~\ref{sec:latent}. 
At a minimum, we know that ordinary thermal conduction will operate regardless of other fluid dynamic processes that may additionally contribute to a Jovian planet's heat flow across a stable helium rain layer. 
For a given internal heat flux, the temperature lapse rate depends on the local thermal conductivity $k$
\begin{equation}
\frac{dT}{dr} = \frac{F}{k}
\end{equation}
where $F$ is the heat flux; the above equation is sometimes called Fourier's Law. 
Estimates for thermal conductivity for a hydrogen-helium mixture under the relevant thermodynamic conditions range from $\sim 20 - 300$W/(m~K) \citep{stevenson-salpeter1977a, french+2012, militzer+2016, becker+2018, Preising+2023}, consistent to about an order of magnitude. 
For our purposes, we will adopt the more recent estimates of 300W/(m~K). 
As an \change{approximate estimate of the luminosity accommodated across the helium rain layer,} we assume the heat flux across the helium rain layer should be on the same order of magnitude as the planet's total inernal heat flux, around 7.5 and 2.9~W/m$^2$ for Jupiter and Saturn respectively \citep{li+2018, wang+2024}.
Using published estimates of thermal conductivity, the corresponding temperature gradient is $dT/dr \sim 25$K/km for Jupiter and $\sim 10$K/km for Saturn. 
From hydrostatic equilibrium, we can rewrite Fourier's Law using pressure instead of radius as the independent variable as 
\begin{equation}
\frac{dT}{dp} = \frac{F}{\rho g k}.
\end{equation}
Using the above equation, we show some examples of possible stable layer profiles in Fig.~\ref{fig:lapses} for Jupiter as envisioned by this work (red curves), compared to an example interior model of Jupiter that treats the boundaries of the helium rain layer as free parameters in order to fit Jupiter's observed gravity harmonics \citep[][blue dashed curve]{militzer+2022}. 

The top of the stable layer is determined by the intersection of the envelope's pressure-temperature (PT) structure with the phase boundary curve corresponding to the envelope's composition (see contours of constant composition in Fig.~\ref{fig:master}). 
For our calculations (red curves in Fig.~\ref{fig:lapses}), we assume the PT structure of the envelope to be isentropic until this intersection point. 
The temperature structure is then permitted to be super-adiabatic (Sec.~\ref{sec:stability}), and will remain so until it intersects the phase boundary corresponding to the helium-enriched deep interior. 
The solid red curve in Fig.~\ref{fig:lapses} shows the lower bound for stable layer thickness using Fourier's Law for conductive heat transport (HT), Jupiter's heat flux chosen to be its global value of 7.5W/m$^2$, and a thermal conductivity of 300W/(m~K). \\

\change{For these values of the heat flux and conductivity, the stable region should be razor-sharp: Assuming $\Delta T\sim 500$\,K  for Jupiter and 1000\,K for Saturn, we obtain a thickness of only 20 and 100\,km, respectively---both very thin compared to prior models.  
We stress that the fact that $\alpha_{\rm eff}<0$ in this region ensures that it is not directly unstable: the density still increases with depth, due to the strong increase in helium abundance. 
However, other advective flows, for example those arising from breaking gravity waves, may occur and lead to additional heat and molecular transport, increasing that thickness. 
Fig.~\ref{fig:lapses} thus shows examples of less-steep} temperature gradients within the stable layer. 
These curved, labeled enhanced in the figure, follow the same procedure but assume thermal conductivity that is enhanced by an order of magnitude (dashed red curve) and a factor of 20 (dotted red curve) to crudely model the behavior of possibly more efficient heat transport mechanisms like turbulent diffusion. 
Fig.~\ref{fig:lapses} shows that shallower temperature lapse rates correspond to cooler temperatures in the helium-rich interior, in agreement with prior studies \citep[e.g.][]{mankovich+2016, mankovich-fortney2020}. \\

\change{We now comment on how the phase diagram behavior dictates the stable layer placement and properties. 
Referring to Fig.~\ref{fig:problem}, we see there exists ambiguity and disagreement about the precise behavior of the H-He phase boundary. 
In order to emulate a variety of behaviors, we follow \cite{mankovich-fortney2020} to modify the dense phase diagram from \cite{schottler-redmer2018} with simple translation. 
While \cite{mankovich-fortney2020} considered translation in temperature $\delta T$, we also consider translation in pressure space $\delta p$ to capture the behavior of phase diagrams that predict the slope of the predict inflection points in the phase boundary at lower pressures 
\citep[see Fig.~\ref{fig:problem} and][]{lorenzen+2011, morales+2013, brygoo+2021}. 
Although these translations affect the predicted depth of the stable layer (Fig.~\ref{fig:lapses}, the stable layer properties as described in this work are robust to a variety of possibilities (see Appendix~\ref{sec:phase-diagram} for further discussion). 
Thus the goal of empirically constraining the depth of the helium rain layer may itself empirically constrain the properties of the H-He phase diagram; such a result would therefore be important for both planetary science and materials physics.
}\\

Based on the temperature gradient derived from the assumption of conductive equilibrium, we can estimate the Brunt--V\"{a}is\"{a}l\"{a} frequency associated with the stable layer. 
As we have argued, the temperature gradient should be sufficiently steep that we can use the approximation that $\frac{d \ln T}{dz} \gg \frac{d \ln p}{dz}$. 
Then we calculate 
\begin{equation}
N^2 = \frac{g}{\rho} \frac{\partial \rho}{\partial T} \left[ \frac{dT}{dr} - \left(\frac{\partial T}{\partial r} \right)_{\rm ad} \right] \sim \alpha_{\rm eff} g \left[ \frac{dT}{dr} - \left(\frac{\partial T}{\partial r}\right)_{\rm ad} \right].
\end{equation}
From Fig.~\ref{fig:master}, we see $\alpha_{\rm eff} \sim -10^{-5}$ in the region of interest. 
Then using a temperature lapse rate of 10K/km and assuming $\frac{dT}{dr} \gg \left(\frac{\partial T}{\partial r}\right)_{\rm ad}$ , we obtain $N \sim 10^{-3}\text{s}^{-1}$ for both Jupiter and Saturn to within an order of magnitude, corresponding to a period on the order of an hour. 
This estimate for $N$ is an upper bound assuming the maximum temperature lapse rate that corresponds to conductive equilibrium. 
Less steep lapse rates facilitated by non-conductive heat transfer processes (for example eddy diffusion through turbulence induced by the breaking of gravity waves as is observed in the Earth's stratosphere) are permitted, and would correspond to a smaller value for $N$. 
The upper bound Brunt--V\"{a}is\"{a}l\"{a} frequency of $N \sim 10^{-3}\text{s}^{-1}$ is near the expected frequencies of the fundamental normal modes on Jupiter and Saturn \citep[see e.g.,][]{markham-stevenson2018}, so considering the helium rain layer may be important for understanding some normal mode eigenfrequencies \citep{fuller2014, mankovich-fuller2021}. 
An estimate for $N$ further allows us to verify the stability of the helium rain layer to diffusive convection, the subject of Sec.~\ref{sec:condensation}. 

\subsection{The effect of latent heat on stable layer thickness}
\label{sec:latent}
\cite{leconte++2024} simulated atmospheres under conditions where convective inhibition was expected from linear theory \citep[e.g.,][]{guillot1995, leconte+2017}. 
While linear theory had previously emphasized  the likely steady-state to be radiative-convective equilibrium \citep[mathematically equivalent to conductive-convective equilibrium in the optically thick case, although subtly different in the optically thin case, see][]{markham-stevenson2021}, \cite{leconte++2024} actually found the stable layer to be thicker in simulations than predicted by theory. 
The difference arose primarily because of the role of latent heat transport from raindrops falling through the stable layer, contributing a non-negligible component to the total planetary heat flux and reducing the temperature gradient required to accommodate the remaining flux. 
Guided by this result, we seek to estimate the importance of the latent heat of helium raindrops on the overall heat flow of Jovian planets, and its possible effects on the thickness of the stable layer. 
\\

We begin by estimating the latent heat associated with the formation of helium-rich raindrops out of a hydrogen-dominated H/He mixture. 
Following \cite{landau-lifshitz_statmech} Chapter 8, Eq.~89.6, we can write the heat released by precipitation of a unit mass helium raindrop out of a solution of helium in hydrogen as 
\begin{equation}
\lambda = \frac{k_B T}{\mu} \left(\frac{\partial \ln y}{\partial \ln T} \right)_{\rm ph},
\label{eq:lambda}
\end{equation}
where the $\rm ph$ subscript refers to the phase boundary at constant pressure as shown in Fig.~\ref{fig:coexistence}. 
The derivation of Eq.~\ref{eq:lambda} requires the solution to be weak, $y\ll 1$. 
In the context of helium separation out of giant planets, we are interested in the regime $y \le y_\odot \sim 8\% \ll 1$, so this assumption is not too unreasonable for our order-of-magnitude interests. 
Additionally, Eq.~\ref{eq:lambda} makes the assumption that the solute (in our case helium) is either dissolved in the solvent (hydrogen) or exists as a pure, undissolved substance. 
In our case, we face the complication that helium raindrops are impure, containing hydrogen impurity $(1-y)\sim20\%$ according to Fig.~\ref{fig:coexistence}. 
Nevertheless although Eq.~\ref{eq:lambda} may be inexact, and it is worthwhile for future research in condensed matter physics to more closely inspect the question of latent heat directly, we consider our estimate sufficient given the limited available information, especially considering the uncertainties about the phase diagram itself (for further discussion, see Appendix~\ref{sec:phase-diagram}). \\

From Fig.~\ref{fig:coexistence}, we see that the solubility of helium in hydrogen increases with increasing temperature, implying that the precipitation of helium raindrops must be an exothermic process, so we are dealing with a positive latent heat. 
The steepness of the dependence of saturation concentration on temperature is then related to the latent heat according to Eq.~\ref{eq:lambda} so long as one has information about the phase diagram. 
\begin{figure}
\centering
\includegraphics[width=0.6\linewidth]{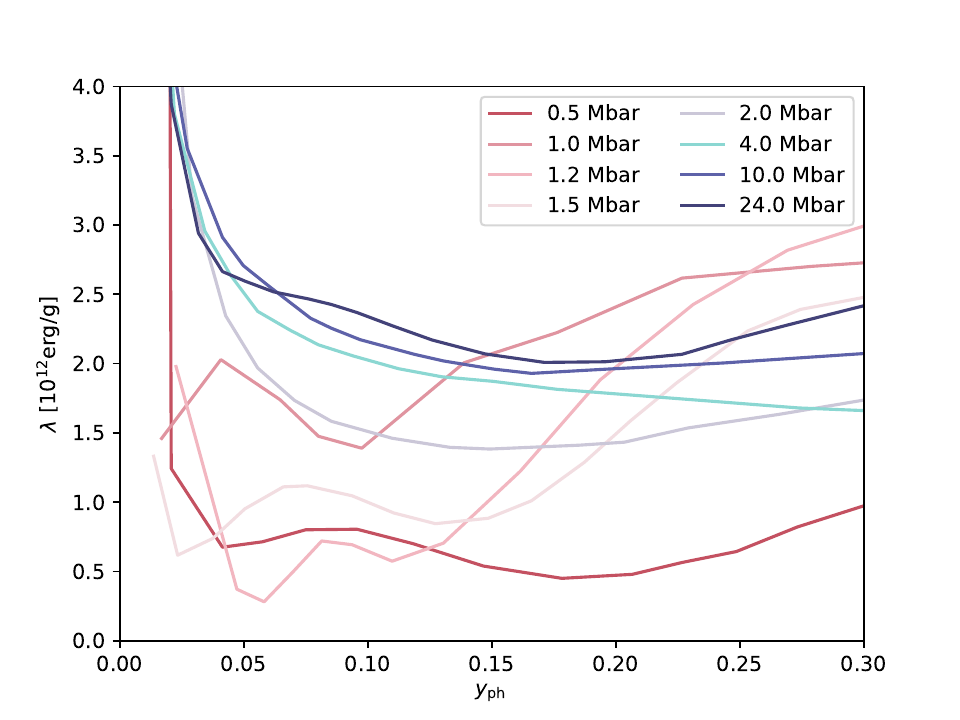}
\caption{Latent heat associated with the formation of helium raindrops at different pressure levels and envelope helium concentration, using data from \cite{schottler-redmer2018} (Fig.~\ref{fig:coexistence}) and Eq.~\ref{eq:lambda}.}
\label{fig:latent}
\end{figure}
We plot our results on Fig.~\ref{fig:latent}. 
\change{We plot our results on Fig. 5. Apart from more undetermined region for very low helium concentrations ($y\lesssim 0.04$) which are generally not expected in the context of Jupiter and Saturn, the values range between 0.5 and $2.5\times 10^{12}$\,erg/g, with variations that are certainly at least partially due to numerical uncertainties in the table interpolations and derivations. For the range of parameters that we consider, i.e., pressures of a few Mbar and values of $y> 0.05$, $\lambda = 2 \times 10^{12}$\,erg/g is a convenient value that we adopt. We note that it corresponds to 200,000\,J/g, almost two orders of magnitude larger than water’s latent heat for vaporization. Heuristically, we note that this may be due both to the order of magnitude increase in temperatures and to the order of magnitude decrease in mean molecular weight for the hydrogen-helium mixture, so that $\beta = L\mu/RT \sim 50$ for $T\sim 6000\,$K, not far from water’s $\beta=19.86$ at $0^\circ$C and other common atmospheric condensing species \citep[see][]{guillot1995}. Obviously, a more detailed calculation, using ab-initio simulations of the multi-component system is warranted. }\\

We now seek an order-of-magnitude estimate for the importance of the latent heat flux from helium rain for the thermal histories of Jupiter and Saturn. 
We do not attempt a detailed evolutionary calculation, but instead seek an approximate luminosity using 
\begin{equation}
    L \sim \frac{\lambda \Delta Y M_{\rm env}}{\Delta t}
\end{equation}
where $\Delta Y$ is the depletion in mass concentration of the planet's contemporary envelope relative to protosolar, $M_{\rm envelope}$ is the mass of the helium-poor envelope above the helium rain later, and $\Delta t$ is the time since the initation of phase separation (IPS). 
We begin with Jupiter. 
We assume H/He phase separation began in the past $\sim$Gyr, and that only about 12\% of Jupiter's mass lies in the helium-depleted envelope. 
From Galileo measurements, we know Jupiter's mass concentration of helium $\Delta Y$ has changed by only about 0.04 relative to the protosolar value. 
Plugging in numbers for Jupiter, we get $L \sim 5.52 \times 10^{23}$erg/s, about an order-of-magnitude smaller than its present-day observed luminosity of $4.61 \times 10^{24}$erg/s. 
If we repeat the calculation for Saturn, we note an important difference: a significantly larger fraction of Saturn's mass lies above the expected helium rain layer, meaning a correspondingly larger flux of helium raindrops must cross the helium rain layer in order for that helium to be removed from Saturn's envelope. 
Assuming $\Delta Y = 0.18$ and $M_{\rm env} = 0.38$, and allowing for $\Delta t\sim 4$Gyr, the same calculation for Saturn yields $L \sim 6.2\times 10^{23}$erg/s, the same order of magnitude as its presently observed luminosity of $8.63\times 10^{23}$erg/s. 

\change{
This implies that conduction would have to transport a much smaller heat flux, leading to a temperature gradient that is proportionally smaller and therefore a thicker helium rain region. Given the uncertainties and the fact that deep inside Saturn’s interior, the luminosity is expected to be smaller than at the surface \citep{mankovich+2016}, it is conceivable that the remaining conductive flux is small, opening the possibility that the helium rain region extends all the way down into the deep regions as expected by present-day traditional models of this planet \citep[e.g.][]{mankovich-fuller2021, howard++2024}. }

\subsection{Implications for internal structure and evolution models}
\change{
In the presence of a stable stratification of the helium rain layer, the temperature structure is defined by (1) which fraction of the total luminosity is transported locally by phenomenons such as wave propagation or droplet rainout and subsequent evaporation, and by (2) how efficiently conduction and radiation may transport that fraction of the luminosity, i.e., the local conductivity/opacity. 
Thus, latent heat release from evaporating drops is not providing any extra gain or loss of energy but only a means to transport that energy throughout the helium rain region. 
We also note that while the moist adiabatic temperature gradient in the helium rain region is of course affected by latent heat (see Appendix~\ref{sec:pseudo}), this is expected to have little consequence for the temperature of that region, given that it is not set by convective processes.  }\\

\change{
Several works have attempted to examine how the interior structure and evolution of Jupiter and Saturn changes in the presence of a super-adiabatic helium-rain layer \citep[for some recent examples, see e.g.,][]{nettelmann++2015, mankovich+2016, mankovich-fortney2020, howard++2024}. 
These works were envisioned in the framework of a double-diffusive instability, with a parameter \citep[$R_\rho$, see][]{garaud2013} controlling the level of super-adiabaticity in the region. These models generally lead to solutions in which most of the increase in helium abundance is pushed to the deeper levels. 
We stress that with a zone that becomes stable even against double-diffusive processes, as we argue in Sec.~\ref{sec:condensation} the superadiabaticity becomes independent of the mean molecular weight gradient. Therefore, unless the latent heat flux is large, superadiabaticity should occur at shallower pressure levels than appreciated in prior works, almost as soon as phase separation begins. }\\

\change{
These considerations have consequences for interior models and how they may reproduce both Juno and Cassini data and for evolution models. 
In the case of Jupiter, we observe that while some interior models that fit Juno gravity data include an extended helium-rain region \citep{militzer+2022}, others are constructed with a discontinuity, where a parameter defines the pressure at which the helium-poor to helium-rich transition occurs \citep{howard+2023}. These models also successfully reproduce the gravitational constraints, with the transition pressure generally falling between 1.5 and 4 Mbar \citep[see figs. C.1 to C.5 in][]{howard+2023}. The presence of a relatively low-pressure thin helium-rain region is therefore compatible with the gravitational constraints, even though the problem of the low-metallicity envelope remains. Similarly, for Jupiter’s evolution, the relatively recent occurrence of helium separation and uncertainties on phase diagram and EoS \citep{howard++2024} imply that these models, with proper fitting, should be compatible with the solar system’s age.}\\

\change{
For Saturn, these issues are probably more severe if we were to envision a very thin and low-pressure helium-rain region since the process started early. 
The fact that most of the luminosity may be carried by falling drops alleviates this issue, at least partially, leading to interior that should be close to the classical ones constrained by gravity and seismology \citep{mankovich-fuller2021}. The question of its evolution \citep[e.g.][]{fortney+hubbard2003, mankovich+2016, mankovich-fortney2020, howard++2024} however requires dedicated calculations.}

\subsection{Thickness of helium rain layers and external magnetic fields}
\label{sec:magnetic}

\begin{figure}
\centering
\includegraphics[width=\linewidth]{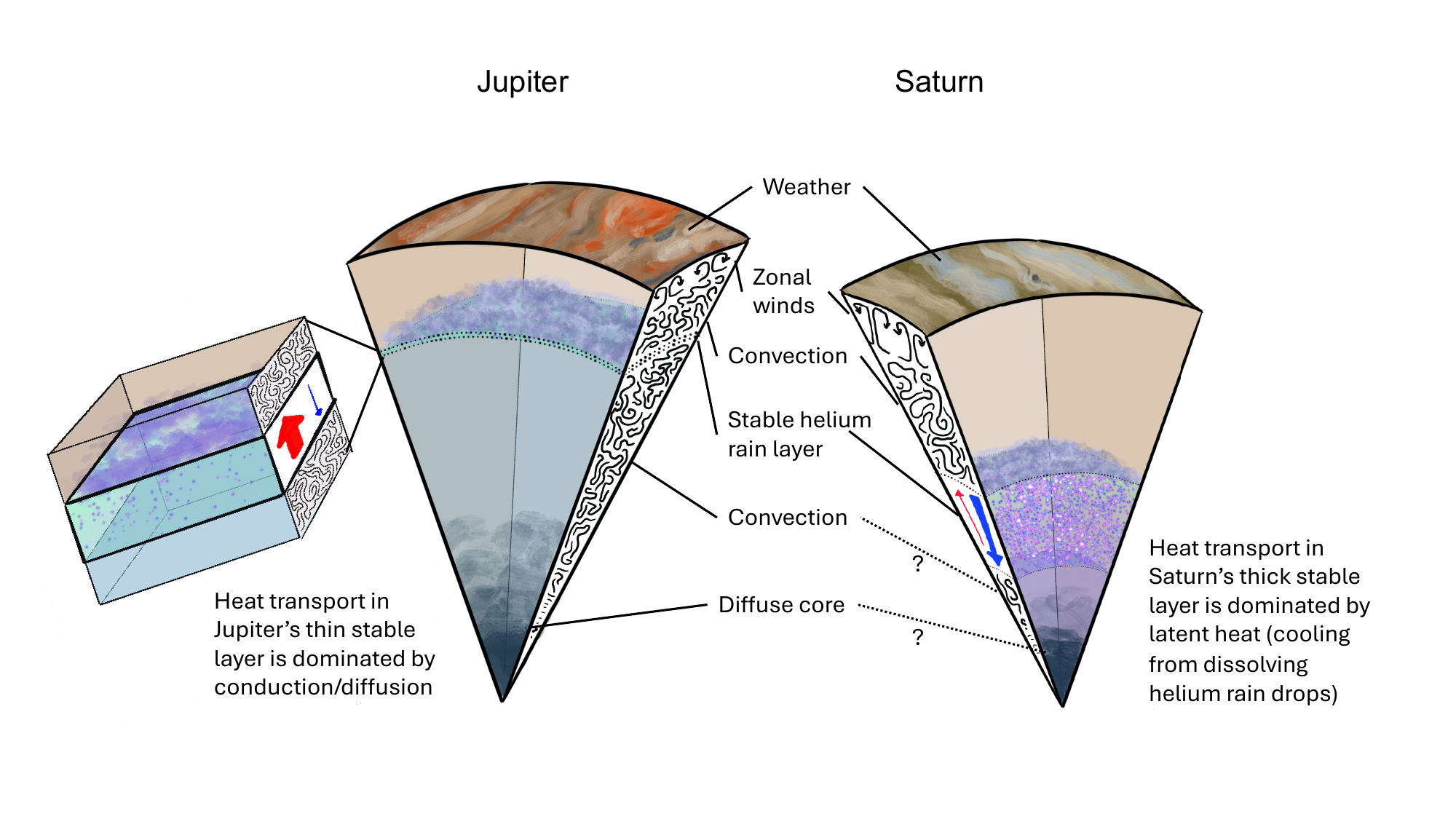}
\caption{Comparison of our proposed helium rain layer structures in the interior of Jupiter and Saturn. We argue Jupiter's helium rain layer may be thin in order for its internal heat flux to be accommodated by diffusion, while Saturn's may be thick because its internal heat flux may be accommodated primarily by latent heat of helium raindrops.} 
\label{fig:cartoon}
\end{figure}

Although these calculations are extremely simplified, they sketch a scenario that appears to bring together several constraints of the problem: The constraints from both Jupiter and Saturn's gravity fields and evolution, the seismological constraints for Saturn and the fact that both planet have extremely different magnetic fields. 
Jupiter's greater mass causes its interior to reach Mbar pressure levels at shallower depths relative to Saturn.  
Therefore a smaller fraction of Jupiter's total mass lies above its helium rain layer. 
Thus depleting Jupiter's molecular envelope by some increment of helium concentration requires relatively less helium rain mass flux to traverse Jupiter's stable helium rain layer compared to Saturn's. 
Therefore on Jupiter, the latent heat transported by settling helium raindrops is insufficient to accommodate its total heat flux, and heat must be transported by diffusion, \change{leading to a large temperature gradient and a thin helium-rain layer. Saturn’s much larger helium rain flux leads to the possibility that raindrops carry all of the heat flux locally, leading to an extended helium rain region.}
We sketch the proposed differences in Jupiter and Saturn's internal structures on Fig.~\ref{fig:cartoon}. \\

\change{Jupiter’s thin helium rain layer is sandwiched between two vigorously convective zones. Contrary to conventional models, the bottom of the helium rain layer now plausibly occurs between 80 and 85\% of Jupiter’s radius, i.e., within the allowed range for the Lowes radius estimates \citep{connerney+2022, sharan+2022}. Even though Jupiter’s top of the dynamo region may be more difficult to define and link to the Lowes estimate than the Earth’s, the fact that Jupiter’s helium rain layer bottom aligns with these estimates provides an interesting consistency between observational data and model expectations.}

\change{On the other hand, Saturn’s interior is characterized by an extended convective but mostly insulating or low-conductivity envelope and a deep stable helium-rain region. Convection in the region below might lead to a dynamo that is both weaker and more dipolar than Jupiter as shown by Cassini's measurements \citep{dougherty+2018}, but the question of its extreme axisymmetry remains.}


\cite{stevenson1982} posited that the differences between Jupiter and Saturn's magnetic field, specifically that Saturn's magnetic field is axisymmetric and precisely alignment with Saturn's spin pole while Jupiter's is not, can be explained by Saturn possessing a stable helium rain layer while Jupiter does not. 
Since the Galileo probe measured sub-solar helium abundance in Jupiter's atmosphere \citep{niemann++1996}, we now believe Jupiter to likewise possess a helium rain layer \citep{young2003}. 
Indeed, a stable layer somewhere in Jupiter's interior is favored by dynamo models that reproduce Jupiter's observed magnetic field \citep{moore+2022}. 
Our findings are thus broadly qualitatively consistent with observations of Saturn's 
luminosity, and magnetic field; as well as Jupiter's magnetic field. \\

Further analysis of Jupiter and Saturn's magnetic fields may prove incisive for further constraining the thickness of the stable layers in Jupiter and Saturn. 
While we place a theoretical lower bound in Sec.~\ref{sec:conductive}, observational constraints are needed. 
\cite{stevenson1982} found that an electrically conductive layer of static stability rotating relative to an underlying magnetic field will attenuate non-axisymmetric components of the field. 
The attenuation factor scales strongly with the magnetic Reynolds number and stable layer thickness. 
If the non-axisymmetric components of Jupiter's magnetic field can be convincingly demonstrated to be systematically weaker that its axisymmetric components, this would both demonstrate that such a mechanism is likely to be at play, and can further observationally constrain the stable layer thickness (albeit with some degeneracy between stable layer thickness and magnetic Reynolds number). 
Saturn's magnetic field can plausibly be understood in a similar way, with a thick stable layer able to efficiently attenuate non-axisymmetric components of the underlying magnetic field generated by a dynamo. \\

However, there are additional complications for both planets that should be investigated further in future work. 
For one thing, based on Fig.~\ref{fig:problem}, the fluid above the stable helium rain layer on both Jupiter and Saturn \change{have} non-negligible electrical conductivity, and is therefore capable of dynamo action. 
Furthermore, settling helium raindrops within the stable layer of Saturn may plausibly generate local fluid motions that themselves could be capable of some dynamo action. 
Careful modeling and data analysis of Jupiter and Saturn's detailed magnetic field structures therefore offer a promising path toward observational constraints on the structure and dynamics of Jupiter and Saturn's helium rain layers. 

\section{Conclusion}
\label{sec:conclusion}
In this work, we have established that the phase separation of hydrogen and helium can inhibit convection under conditions relevant to the interiors of Jovian planets. 
The most important specific conclusions are as follows: 
\begin{enumerate}
\item The effective coefficient of thermal expansion $\alpha_{\rm eff}$ of hydrogen-helium mixtures in the phase-separating region of parameter space is negative, implying that convection should be inhibited\change{, even for super-adiabatic temperature gradients within the helium rain layer. }
\item The lower limit for stable layer thickness in conductive equilibrium is thin (20--100km on Jupiter and Saturn respectively), although thicker stable regions are permitted if additional thermal transport mechanisms are more efficient than conduction.
\item With a latent heat of helium raindrops estimated at $\sim 2 \times 10^{12}\,$erg/g \change{(with a significant uncertainty)}, the latent heat flux associated with helium rain is sufficient to accommodate nearly Saturn's entire luminosity, but not Jupiter's. This could allow Saturn to possess a thick helium rain layer even while Jupiter's is thin.
\item The stable helium rain layer as described in this work is consistent with studies of Jupiter and Saturn's magnetic fields that require a region of static stability near the helium rain layer. Further analysis of Jupiter and Saturn's magnetic field spectrum, especially comparing axisymmetric to non-axisymmetric harmonic strength, may offer observational constraints on their respective stable layer thicknesses.
\item The region of static stability can host \change{internal gravity waves that} may affect Jupiter and Saturn's seismic spectra.
\end{enumerate}

Fruitful topics for future work involve the interplay between the stability of the helium rain layer and the generation of Jupiter's magnetic field. 
Further analysis of Jupiter's magnetic field spectrum are warranted, for example to determine whether there exists an attenuation of the non-axisymmetric components of Jupiter's magnetic field. 

Further understanding the structure of the helium rain layer requires continued computational and experimental study of the physical properties of hydrogen-helium mixtures under Jovian interior conditions. 
The field of giant planet interior science would therefore benefit from greater investment into DFT simulations, and especially high pressure experiments.

\subsection*{Acknowledgements}
This work has been funded by CNES, the CNES postdoctoral program, and the Clyde and Patricia Tombaugh Scholar fund. 
We would like to especially thank Saburo Howard for providing tabulated data for the EoS calculations, along with detailed conversations and feedback about planetary evolution models. 
We would also like to acknowledge Dave Stevenson, Jack Connerney, Hao Cao, and Jeremy Bloxham for valuable discussions about interpreting Jupiter's magnetic field, Ronald Redmer for advice on equations of state and hydrogen-helium mixtures, Chris Mankovich for useful discussions about Saturn's evolution, and the anonymous peer reviewers for their constructive feedback.

\appendix
The main text refers to a host of detailed issues, such as whether the uncertainties in the H/He phase diagram affect our central conclusions, the relationship between envelope cooling and helium depletion, the pseudo-adiabatic lapse rate referenced in Eq.~\ref{eq:instability-criterion}, and the compatibility of our arguments about the internal temperature of Jovian planets with energy conservation.
In the main text, we aimed to make our arguments as concise and straightforward as possible for clarity and readability. 
However, some readers may be interested in or skeptical of some of these details. 
We therefore provide substantially expanded discussion of these topics below. 

In the main text, we considered the phase diagram of a hydrogen-helium mixture under conditions relevant to the interiors of Jovian planets. 
For one thing, the hydrogen-helium phase diagram is currently uncertain, and we must consider how alterations to the phase diagram may affect our conclusions. 
This will be the focus of Sec.~\ref{sec:phase-diagram}. 
Sec.~\ref{sec:degeneracy} \change{considers the degeneracy between allowed phase-boundary crossings and comments on the ambiguity this may introduce with respect to future thermal evolution modeling.} 
Finally, the main text discusses adiabats in the conventional dry adiabatic sense. 
However, we know from experience in Earth's atmosphere that adiabats are modified when material undergoes a first order phase transition. 
We therefore comment on this modification in Sec.~\ref{sec:pseudo}. \\
 
\section{Uncertainties in the hydrogen-helium phase diagram and equation of state}
\label{sec:phase-diagram}
Although the theory of hydrogen-helium phase separation is not controversial, the details are currently poorly constrained. 
The current state-of-the-art for our understanding of the phase diagram for the coexistence of hydrogen and helium employ both experimental and computational methods. 
Computational methods have the advantage of opening the possibility of more fully exploring phase space, for example using functional theory molecular dynamics simulations (DFT-MD), as has been done using van der Waals \citep{schottler-redmer2018} and Perdew-Burke-Ernzerhof \citep{lorenzen+2011, morales+2013} exchange-correlation functionals. 
The two approaches agree that phase separation occurs, but the shape of the phase diagram is substantially different.  
Moreover, neither result is consistent with claimed experimental observations of hydrogen-helium phase separation \citep{brygoo+2021}. 
Fig.~\ref{fig:problem} shows the relationship between each of these predictions for the phase boundary between fully mixed and partly de-mixed hydrogen and helium for a protosolar helium mixing ratio superimposed on planetary adiabats for the interiors of Jupiter and Saturn. 
These different predictions for the hydrogen-helium phase diagram have interesting implications for the structure and evolution of Jovian planets, but those details are beyond the scope of this work \citep[for pertinent discussion, see][]{mankovich-fortney2020, howard++2024}. \\

Whether different predictions for the phase diagram affect our result from Sec.~\ref{sec:alpha} depends on the detailed behavior of the phase diagram, but we can comment on the direction and magnitude of the change under various conditions. 
\cite{mankovich-fortney2020} considered modifying the phase diagram by simple translation, so that $y_1(T) = y_0(T - \delta T)$ where $y_0(T)$ refers to results from \cite{schottler-redmer2018} and $y_1(T)$ refers to the modified equivalent phase curve. 
In that work, $\delta T$ was left as a free parameter. 
\change{In this work, we generalize permitted translations to existing phase diagrams so that offsets in pressure space are also permitted, so that phase boundary curves become $y_1(p,T) = y_0(p+\delta p, T-\delta T)$. }
For the subject of interest for this work, we estimate the importance of such a modifications. 
From Equation~\ref{eq:alpha_eff}, $\frac{dy}{dT}$ and $y$ evaluated at $T+\delta T$ would be unchanged. 
The modification would be to the various other terms in the expression which would need to be evaluated at $T+\delta T$. 
We experimented with changing the phase diagram in this way, and found it had practically no effect on the central conclusion that convection should be strongly inhibited; we continue to find $\alpha_{\rm eff} < 0$ virtually everywhere. 
We can rationalize this result intuitively by inspecting each term in Equation~\ref{eq:alpha_eff}. 
Using the method in Sec.~\ref{sec:alpha}, we find $\alpha$ to be a weak function of $T$, and the magnitude is generally about an order of magnitude smaller than the second right hand side term (indeed, this is precise reason that $\alpha_{\rm eff} < 0$ throughout most of parameter space). 
Therefore we do not expect evaluating $\alpha$ at a different temperature to drastically effect the sign of $\alpha_{\rm eff}$. 
We then compare the effect intermolecular spacing $\frac{1}{V} \left( \frac{\partial V}{\partial y} \right)$ to the effect on the change of mean molecular weight $\mu \left(\frac{\partial (1/\mu)}{\partial y}\right)$, an expansion of the contributions to the bracketed terms on the right hand side of Equation~\ref{eq:alpha_eff}. 
The first term is a subtle effect due to changes in molecular size and spacing, and is of order a few percent \citep{howard-guillot2023}. 
The second term is a major effect due to changing composition as condensation occurs, and is of order unity for a hydrogen-helium mixture \citep[for further discussion, see][]{markham+2022}. 
\change{Reasonable translations in pressure space are likewise relatively unimportant, as the coefficient of thermal expansion of H-He mixtures is also a relatively weak function of pressure in the pressure range of interest ($\sim 1$--a few Mbar). }
From these arguments, simple translation of the phase separation curve has only a marginal effect on our results. \\

Another way to rationalize the differences in the phase boundary curves from Fig.~\ref{fig:problem} may be transformed in more general ways than simple translation. 
In this case, instead of each curve in Fig.~\ref{fig:coexistence} being translated by a fixed quantity $\delta T$ at each pressure level, we could instead imagine stretching them such that they possess a much higher critical temperature, but a comparable condensation temperature at a given pressure. 
Such a transformation would increase $\frac{dT}{dy}$ along the phase boundary, or equivalently decrease $\frac{dy}{dT}$. 
If $\frac{dy}{dT}$ were decreased sufficiently, it could diminish the magnitude of the second right hand side term in Equation~\ref{eq:alpha_eff}. 
Therefore such a transformation would be far more impactful than a simple translation. 
As an example of one of more extreme cases, we can take the 1.5Mbar curve from Fig.~\ref{fig:coexistence} and retain the temperature at $y=0$, but increase the temperature at $y=8.87\%$ by 6000K based on the difference between the \cite{schottler-redmer2018} and \cite{brygoo+2021} curves in Fig.~\ref{fig:problem}. 
Doing so would decrease the characteristic magnitude of $\frac{dy}{dT}$ by about a factor of three. 
In this case, we may not expect convection to be inhibited everywhere. \\

In reality, our understanding of the hydrogen-helium phase diagram remains imperfect. 
However, if the behavior approximately resembles the calculations from \cite{schottler-redmer2018}, then we expect convective inhibition to be a robust result. 
The fact that magnetic field measurements for both Jupiter and Saturn appear to be consistent with the existence of a stably stratified layer near the hydrogen-helium de-mixing level may be interpreted as evidence that the basic shape of the phase diagram from \cite{schottler-redmer2018} may be plausible, even if some modifications may be necessary \citep{mankovich-fortney2020, howard++2024}. 

\section{Non-uniqueness of phase boundary crossings}
\label{sec:degeneracy}
Throughout this work, we have discussed the intersection between a given adiabat and a given phase boundary to define the top of the helium rain layer. 
Doing so implies assuming a priori a helium abundance $y$. 
Then for an adiabat with a specified entropy (or equivalently a specified 1 bar temperature), one can draw a unique phase curve using Fig.~\ref{fig:master}. 
Thus one infers a unique intersection point. 
While this methodology appears straightforward initially, there are a variety of complications swept under the rug in the main text that we discuss in greater detail here. \\

\begin{figure}
\centering
\includegraphics[width=0.5\textwidth]{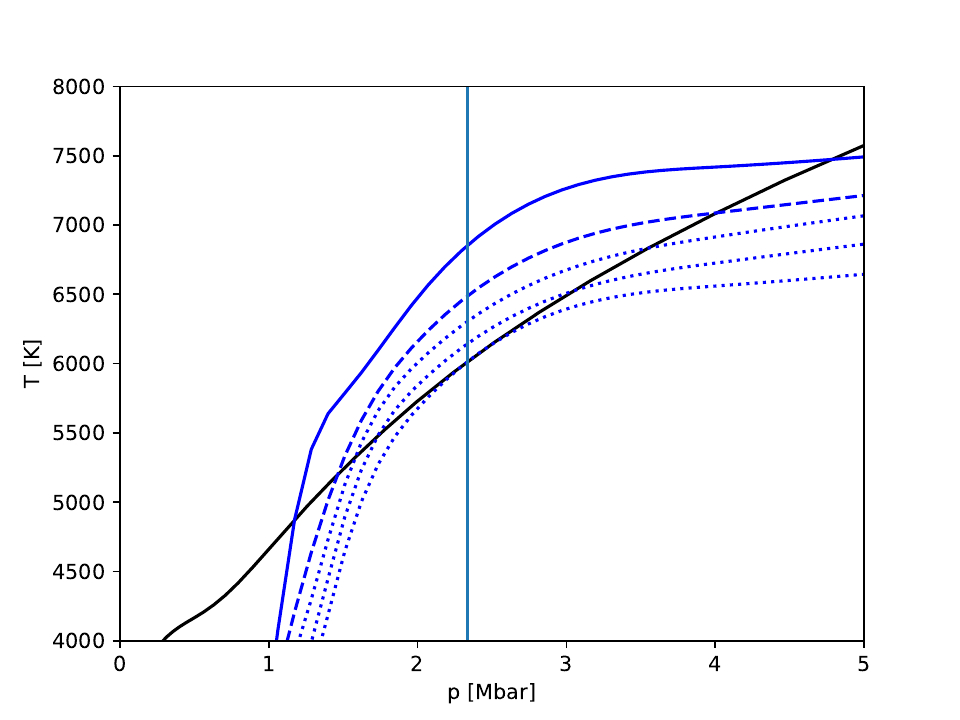}
\caption{Figure illustrating the non-uniqueness of intersection points for an arbitrary adiabat (solid black). The adiabat intersects the proto-solar phase boundary (solid blue), the present Jovian phase boundary (dashed blue), and a variety of phase boundaries further depleted in helium than Jupiter's present envelope (dotted blue). The top of the helium rain level corresponding to maximum rainout is shown as a vertical line, where the adiabat has a common tangent point with its corresponding phase boundary curve. }
\label{fig:degeneracy}
\end{figure}

In this section we set aside the issue of phase diagram uncertainty discussed in Sec.~\ref{sec:phase-diagram} to illustrate a different problem. 
Even if the phase diagram were known with precise certainty, the relationship between envelope 1 bar equivalent temperature and helium enrichment remains uncertain. 
Intuitively, as the planet cools and intersects the phase diagram after IPS, phase separation will occur and envelope helium depletion will ensue. 
However, on the basis of thermodynamics alone, one cannot decisively conclude the extent to which helium is depleted. 
In reality there does not exist a single adiabatic curve and a single phase boundary curve---the phase diagram specifies a two dimensional phase boundary surface in three dimensional pressure-temperature-composition (PTX) space. 
Meanwhile there likewise exist a continuum of adiabats with identical 1 bar equivalent temperatures but with different compositions which, taken together, constitute a corresponding continuous surface of adiabats in PTX space. 
Therefore there does not exist a single unique intersection point between an adiabat and a phase boundary curve, but rather a continuous one dimensional curve through PTX space of intersections between the surface of adiabats of and the phase boundary surface. 
We illustrate the problem graphically with Fig.~\ref{fig:degeneracy}, which shows how a single adiabat (the PT structure of an adiabat is a weak function of composition, so we treat it as a single curve) intersects with a variety of different phase boundary curves of different compositions. 
The problem is not serious if one can confidently specify the relevant composition, as we can today for Jupiter with the benefit of Galileo measurements. 
However, if one wishes to conduct a thermal evolution calculation and specify composition as a function of 1 bar temperature through the planet's thermal history, or predict how helium abundance will evolve as a function of temperature as evolution proceeds, one must employ additional constraints to choose a unique point from the continuous curve of the intersection of the adiabatic surface and the phase boundary surface. \\
The intersection curve contains an infinite number of points. 
However, it notably does not necessarily extend through the full domain of composition space ($y$ between 0 and 1). 
For some 1 bar temperatures, there exists a minimum possible value of $y$ beneath which there is no intersection point between the adiabat and the phase boundary curve. 
This point corresponds to the common tangent point between the adiabat and the phase boundary, analogous to the IPS point. 
We call this extremum the ``maximum rainout'' point. 
Furthermore, assuming the region above and below the helium rain layer are adiabatic, maximum rainout is the planet's lowest-energy state for a given 1 bar temperature and therefore a natural choice. 
\change{If the maximum-rainout scenario can be more rigorously demonstrated to be the preferred planetary state in future work, then studying Jupiter and Saturn's interiors provides concrete constraints on the hydrogen-helium phase diagram. 
Specifically, if the envelope helium abundance is measured (for example by probe) and the envelope isentrope is specified, then the true hydrogen-helium phase diagram at the specified helium enrichment must possess a common tangent with the isentrope, possibly ruling out certain proposed phase diagrams and providing independent constraints on materials physics.}\\

Different choices are possible. 
For example, one could assume that the helium rain layer pressure level remains constant. 
Picking that pressure level then specifies a point along the curve of intersection between the adiabatic and phase diagram surfaces. 
This method was used by \cite{mankovich-fortney2020}, choosing 2 Mbar (see their Figure 2). 
\change{Again, as shown in Fig.~\ref{fig:degeneracy}, there exists a continuum of physically plausible solutions. 
Future theoretical analysis may provide a more robust physical rationale to prefer a particular solution. } \\

Although one must make some assumptions to specify a unique point, the possible choices are nevertheless partly limited by physics. 
For example, clearly the assumption of zero phase separation is unfeasible. 
The only way for the planet's temperature profile from Fig.~\ref{fig:degeneracy} not to lie below the protosolar phase boundary curve at pressures exceeding $\sim 1.1$ Mbar is to possess a superiadiabatic compositional gradient. 
An interior state with a temperature profile substantially beneath the phase boundary cannot persist for geologic timescales \citep[$\delta T/T \lesssim 1\%$,][]{stevenson-salpeter1977a}. 
In order to avoid this unphysical scenario, there must exist a super-adiabatic temperature gradient that increases the temperature to remain near or above the phase boundary curve. 
In the main text, we argue that such a super-adiabatic temperature gradient is physically permitted, but only insofar as it is accommodated by a corresponding compositional gradient. 
However due to the constraint of mass balance, such a compositional gradient cannot exist without phase separation. 
Therefore the constraint of mass balance hints at the opposite constraint to the maximum-rainout assumption, which we might call minimum-rainout. \\

In addition to the complications discussed in this section, there is an additional complication involving the possibility of layered convection. 
This is not layered convection in the sense of double-diffusive convection that we argued should not operate in the phase-separating region in Sec.~\ref{sec:condensation}. 
However, when one constructing sample interior profiles \change{(e.g., Fig.~\ref{fig:lapses})}, in some cases we find that the adiabatic slope is less than the deep phase boundary slope, indicating that \change{another layer of} phase separation should occur. 
In this case, an additional layer will develop following the same logic as the argument for the first helium rain layer. 
In principle, one can encounter this issue multiple times, and must proceed iteratively until a solution is found where the slope of the adiabat is greater than or equal to the slope of the phase boundary. 
This is a relatively minor issue that does not substantively impact our primary findings, but we include the possibility here in the appendix as it may be of interest to future works interested in details of the dynamics \change{and evolution} of the helium rain layer.

\section{The pseudo-adiabatic lapse rate for hydrogen-helium phase separating mixtures}
\label{sec:pseudo}

\change{We derive hereafter an approximate moist adiabatic gradient relevant for non-ideal deep-interior conditions in a way similar to that obtained in the atmosphere for the condensation of a minor species. A proper adiabat should be calculated directly from a multi-component EOS including non-ideal effects and all phases in presence.}

We define the specific entropy of a multi-component, multi-phase, closed system as
\begin{equation}
s = \sum_i m_i s_i
\label{eq:total-entropy}
\end{equation}
where $m_i$ is the mass fraction of each phase, and $s_i$ is that phase's corresponding specific entropy. 
\change{While Eq.~\ref{eq:total-entropy} does not explicitly contain an entropy of mixing term, entropy of mixing can nevertheless be included implicitly by using an EoS that includes entropy of mixing in its tabulated entropy values \citep[e.g.,][]{howard+2023}. }
By definition, $\sum_i m_i = 1$. 
We write the specific entropy differential as
\begin{equation}
ds = \left(\frac{\partial s}{\partial p}\right)_T dp + \left(\frac{\partial s}{\partial T}\right)_p dT.
\label{eq:ds-general}
\end{equation}
For an arbitrary thermodynamic quantity $q$, we can write using Equation~\ref{eq:total-entropy} 
\begin{equation}
\frac{ds}{dq} = \sum_i \left( m_i \frac{ds_i}{dq} + \frac{dm_i}{dq} s_i \right).
\end{equation}
So for a two phase system, we can rewrite Equation~\ref{eq:ds-general} as
\begin{dmath}
ds = dp\left[ \left(\frac{\partial s_1}{\partial p}\right)_T m_1 + \left(\frac{\partial s_2}{\partial p}\right)_T (1-m_1) + \left(\frac{\partial m_1}{\partial p}\right)_T (s_1 - s_2) \right] + 
dT \left[ \left(\frac{\partial s_1}{\partial T}\right)_p m_1 + \left(\frac{\partial s_2}{\partial T}\right)_p (1-m_1) + \left( \frac{\partial m_1}{\partial T}\right)_p (s_1 - s_2) \right].
\label{eq:ds-long}
\end{dmath}
We now specialize to a system involving two coexisting substances $x$ and $y$ in two coexisting phases 1 and 2. 
For a closed system, the total mass fractions $Y$ and $X = 1-Y$ remains fixed. 
Thus, 
\begin{equation}
Y_1 m_1 + Y_2 (1-m_1) = Y
\end{equation}
\begin{equation}
m_1 = \frac{Y_2 - Y}{Y_2 - Y_1}
\end{equation}
\begin{equation}
\frac{dm_1}{dq} = \frac{1}{(Y_2 - Y_1)^2} \left[\frac{dY_2}{dq}(Y-Y_1) + \frac{dY_1}{dq}(Y_2-Y) \right]
\end{equation}
\begin{equation}
\frac{dm_2}{dq} = - \frac{dm_1}{dq}
\end{equation}
where $Y_i = 1-X_i$ refers to the mass fraction of substance $y$ in phase $i$. 
From the first law of thermodynamics, 
\begin{equation}
du = T ds - p dV + \sum_i g_i dN_i
\end{equation}
where $g_i$ is the specific Gibbs energy of phase $i$ and $V = \mu/\rho$ is the specific volume ($\mu$ is the mean molecular weight). 
\change{In general for coexisting phases in a multicomponent system, one must use data in order to determine the specific Gibbs energy of each phase. 
However, for the special case of simple phase transitions of unitary substances, }
, $g_i$ is equivalent for all phases so that $\sum_i g_i dN_i = 0$. 
\change{We will therefore proceed with the following derivation using the assumption that $\sum_i g_i dN_i = 0$ in order to obtain a clean closed-form analytic result useful for simple phase transitions, although readers are cautioned to account for that term being nonzero if applying these equations to the more general case. }
In the case of \change{a first order phase transition with} no change in internal energy, 
\begin{equation}
\left( \frac{\partial s}{\partial V} \right)_T = \frac{p}{T}.
\end{equation}
We therefore seek to write $\frac{\partial s}{\partial p}$ in terms of $\frac{\partial s}{\partial V}$. 
We can write 
\begin{equation}
\frac{\partial s}{\partial p} = \frac{\partial s}{\partial V} \left(\frac{\partial p}{\partial V}\right)^{-1} = - \frac{p \mu}{\rho K T}
\end{equation}
where $K$ is the isothermal bulk modulus. 
Next we recall from the definition of temperature 
\begin{equation}
\left(\frac{\partial s}{\partial T}\right)_p = \frac{c_p}{T}
\end{equation}
Finally we define the latent heat 
\begin{equation}
\lambda = (s_1 - s_2) T.
\label{eq:latent-heat}
\end{equation}
Now we can rewrite Equation~\ref{eq:ds-long} as 
\begin{dmath}
ds = dT \left[\frac{m_1 c_{p,1}+(1-m_1)c_{p,2}}{T} + \frac{\partial m_1}{\partial T} \frac{\lambda}{T} \right] -dp \left[ \frac{p}{T} \left( \frac{m_1 \mu_1}{\rho_1 K_1} + \frac{(1-m_1)\mu_2}{\rho_2 K_2} \right) - \frac{\partial m_1}{\partial p} \frac{\lambda}{T} \right].
\end{dmath}
Then in the isentropic case $ds=0$, we can write the isentropic lapse rate as 
\begin{equation}
\left(\frac{\partial T}{\partial p}\right)_s = \frac{p \left( \frac{m_1 \mu_1}{\rho_1 K_1} + \frac{(1-m_1)\mu_2}{\rho_2 K_2} \right) - \frac{\partial m_1}{\partial p} \lambda}{m_1 c_{p,1}+(1-m_1)c_{p,2}+ \frac{\partial m_1}{\partial T}\lambda}
\end{equation}
and we are done. \\

To elucidate further, it may be useful to consider the case where $m_2 \ll m_1 \implies m_1 \sim 1$ (sometimes called the pseudo-adiabat in atmospheric contexts). 
Then we can write 
\begin{equation}
\left(\frac{\partial T}{\partial p}\right)_s = \left(\frac{\partial T}{\partial p} \right)_{\rm ad}  \left( \frac{1 - \frac{\lambda \rho K}{\mu p} \frac{\partial m_1}{\partial p} }{1 + \frac{\lambda}{c_p} \frac{\partial m_1}{\partial T} } \right)
\end{equation}
where $\left(\frac{\partial T}{\partial p} \right)_{\rm ad} = \frac{p \mu}{\rho K c_p}$ is the ``dry'' adiabatic lapse rate without phase transitions. 
We can verify for the ideal gas case, $K = p$, $p \mu/\rho = R T$. 
Then $\left( \frac{\partial \ln T}{\partial \ln p} \right)_{\rm ad} = \frac{R}{c_p}$ as expected. 
Similarly the other terms correspond to the classical moist adiabatic lapse rate in the ideal gas limit, and where the partial pressure of the vapor species obeys an Arrhenius relationship \citep[cf.][Chapter 4]{emanuel-book}. 
Thus if one specifies the relevant parameters, one determines the isentropic lapse rate associated with the phase separating region.

\bibliography{library}{}
\bibliographystyle{apalike}

\end{document}